\documentclass[%
 aip,
 jcp,%
 amsmath,amssymb,
 preprint,%
]{revtex4-1}
\usepackage{siunitx}
\usepackage{graphicx}
\usepackage{dcolumn}
\usepackage{bm}
\usepackage{color}
\usepackage{epstopdf}
\usepackage{multirow}
\usepackage[utf8]{inputenc}

\begin{document}


\title{Communication: An efficient algorithm for Cholesky decomposition of electron repulsion integrals}

\author{Sarai D. Folkestad}
\thanks{Contributed equally to this work}
\author{Eirik F. Kj\o nstad}%
\thanks{Contributed equally to this work}
\affiliation{Department of Chemistry, Norwegian University of Science and Technology, N-7491 Trondheim, Norway}%
\affiliation{%
Scuola Normale Superiore, Piazza dei Cavaleri 7, 56126 Pisa, Italy 
}%

\author{Henrik Koch}
\email{henrik.koch@sns.it}
\affiliation{%
Scuola Normale Superiore, Piazza dei Cavaleri 7, 56126 Pisa, Italy 
}%
\affiliation{Department of Chemistry, Norwegian University of Science and Technology, N-7491 Trondheim, Norway}

\date{\today}

\begin{abstract}

We present an algorithm where only the Cholesky basis is determined in the decomposition procedure. This allows for improved screening and a partitioned matrix decomposition scheme, both of which significantly reduce memory usage and computational cost. After the basis has been determined, an inner projection technique is used to construct the Cholesky vectors. The algorithm extends the application range of the methodology and is well suited for multilevel methods. We apply the algorithm to systems with up to 80000 atomic orbitals.

\end{abstract}

\maketitle
\section{Introduction}

The Beebe-Linderberg\cite{Beebe1977} algorithm for the Cholesky decomposition of the electron repulsion integral matrix was developed in the 1970s. Beebe and Linderberg observed that, given the high degree of
linear dependence in the matrix, significant computational savings are obtainable through  decomposition. Furthermore, they identified the approach as an inner projection in the sense introduced by Löwdin.\cite{Lowdin1965,Lowdin1971} 
The algorithm was later modified to include screening by Røeggen and Wisløff-Nilssen,\cite{roeggen1986} who 
also demonstrated that the numerical rank is proportional to the number of atomic orbitals, as had already been suggested.\cite{Beebe1977}
An algorithm suited for large-scale applications was 
first 
proposed in 2003 by Koch, S{\'a}nchez de Mer{\'a}s and Pedersen.\cite{koch2003} This algorithm was implemented in Dalton\cite{Dalton} and 
subsequently
included in the Molcas program.\cite{molcas,Aquilante2008b}
A number of applications based on the Cholesky decomposition of the integrals have since been published.\cite{Pedersen2004,Crawford2005,Aquilante2008c,Aquilante2011}

An inner projection technique introduced by Vahtras, Alml\"of and Feyereisen\cite{Vahtras1993} is often referred to as the resolution of identity (RI) or density fitting method.\cite{Merlot2013} In RI, the inner projection is onto the space spanned by an auxiliary basis.  
 The use of prefitted auxiliary basis sets in this projection is an approach that has gained much popularity.\cite{Eichkorn1995,Eichkorn1997} 
However, while the auxiliary basis in a Cholesky decomposition is systematically improved by lowering the decomposition threshold, this is not the case for prefitted auxiliary basis sets.

One advantage of preoptimized auxiliary bases is that they are usually one-centered, making the integrals at most three-centered and therefore computationally cheaper. A Cholesky basis, on the other hand, typically includes many two-center functions. 
Pedersen and coworkers have advocated the atomic (aCD) and one-center (1C-CD) decomposition methods, where the Cholesky basis is restricted to one-center functions.\cite{Aquilante2007} These methods necessarily imply a certain loss of accuracy.
Nevertheless, the auxiliary basis sets of aCD and 1C-CD are, unlike prefitted bases, not biased toward any method or specific quantity.\cite{Aquilante2007,pedersen2009} 

Alternatively, the computational cost of a Cholesky decomposition may be reduced by controlling the error in method specific quantities, such as the Coulomb or exchange energies, rather than the electron repulsion integrals. 
This type of method specific Cholesky decomposition has been shown to substantially reduce the size of the auxiliary basis with no added loss of accuracy in the target quantities.\cite{Boman2008} The approach is well suited for multilevel methods, where only subsets of integrals are needed in the correlated treatments.\cite{Myhre2014,Myhre2016,Sether2017}

To be generally applicable, an integral approximation scheme must have analytic geometric derivatives. Such derivatives are easily derived for RI using prefitted auxiliary bases.\cite{Hattig2003}
Although not apparent in the early discussion of gradients by O'neal and Simons,\cite{oneal1989} the equivalence of RI and Cholesky decomposition implies that analogous gradient expressions exist for Cholesky decomposed integrals.
Recently, this was exploited to formulate and implement analytic gradients by Aquilante, Lindh and Pedersen.\cite{Aquilante2008}

In this contribution, we introduce an algorithm where only the elements of the auxiliary basis are determined in the decomposition of the matrix.\cite{Alfredo} As a consequence, both the columns and rows of the integral matrix may be screened, giving a reduction in both memory usage and computational cost. Once the basis has been identified, the Cholesky vectors are constructed using the RI formulation of Cholesky decomposition. To illustrate the flexibility of the algorithm, we have 
also implemented 1C-CD,\cite{Aquilante2007} a method specific multilevel screening, and a decomposition scheme using a partitioned integral matrix. 

\section{Theory}
\begin{figure}[tb]
    \centering
    \includegraphics[width=0.45\textwidth]{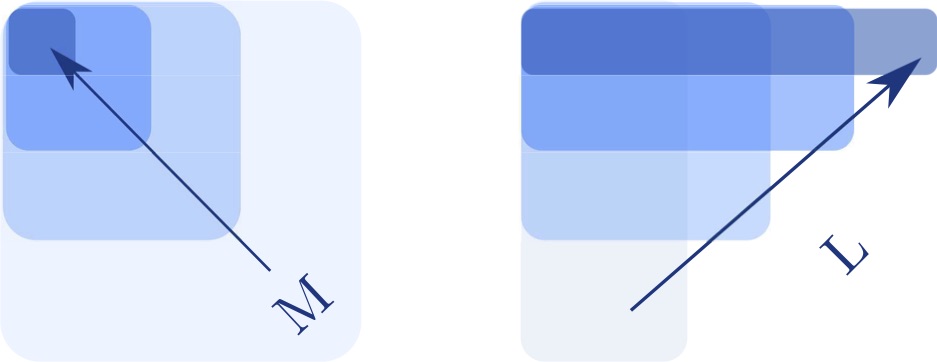}
    \caption{
    While determining the basis $\mathcal{B}$, we screen out elements of $\bm{M}$ and $\bm{L}$ that are no longer needed. This means that we consider $M_{pq}$ and $L_p^{J}$ for $p, q \in \mathcal{D}$ and $J \in \mathcal{B}$. Only $\bm{L}$ is kept in memory throughout the decomposition procedure.
    }
    \label{fig:mem_reduction}
\end{figure}
The electron repulsion integral matrix $\bm{M}$ is symmetric positive semidefinite and may therefore be Cholesky decomposed,
\begin{align}
    M_{\alpha\beta,\gamma\delta} = (\alpha \beta | \gamma \delta) = \sum_{J} L_{\alpha\beta}^{J} L_{\gamma \delta}^{J} = (\bm{L} \bm{L}^T)_{\alpha\beta,\gamma\delta}, \label{eq:CD}
\end{align}
where $\alpha, \beta, \ldots$ denote the real atomic orbitals (AOs) $\{\chi_\alpha (\bm{r}) \}_{\alpha}$. 
Alternatively, $\bm{M}$ may be expressed as an inner projection,
\begin{align}
    M_{\alpha\beta,\gamma\delta} = \sum_{JK} (\alpha \beta | \rho_J)(\bm{S}^{-1})_{JK}(\rho_K| \gamma \delta), \label{eq:CDinner}
\end{align}
where $S_{JK} = (\rho_J | \rho_K)$. The auxiliary functions $\{\rho_J(\bm{r})\}_J$ form a basis for the space spanned by $\{ \chi_\gamma (\bm{r}) \chi_\delta (\bm{r}) \}_{\gamma\delta}$. 
Since $\bm{S} = \bm{Q}\bm{Q}^\mathrm{T}$ for some $\bm{Q}$, we may identify the Cholesky vectors as
\begin{align}
L_{\alpha\beta}^J = \sum_K (\alpha \beta | \rho_K)Q^{-\mathrm{T}}_{KJ}. \label{eq:Cholesky_vec_from_RI_expr}
\end{align}
That is, a Cholesky decomposition is equivalent to an RI approximation.\cite{Beebe1977,Vahtras1993}

In the full-pivoting Cholesky decomposition of $\bm{M}$, one first selects the largest diagonal element $M_{JJ}$ as the pivot. Then, the corresponding Cholesky vector
\begin{align}
    L_p^J = \frac{M_{pJ}}{\sqrt{M_{JJ}}}
\end{align}
is constructed. Finally, $\bm{M}$ is updated according to
\begin{align}
   {M}_{pq} \leftarrow M_{pq} - L^J_p L^J_q.\label{eq:subtract_cholesky}
\end{align}
These steps are repeated until all diagonal elements of $\bm{M}$ are below a given threshold $\tau > 0$. From the Cauchy-Schwarz inequality, 
\begin{align}
    M_{pq}^2 \leq  M_{pp}   M_{qq}, \label{eq:CS}
\end{align}
all elements of $\bm{M}$ will then be smaller than $\tau$ in absolute value. We may thus conclude that
\begin{align}
    M_{pq} \approx \sum_J L_p^J L_q^J,
\end{align}
where the error in $M_{pq}$ is less than $\tau$.

We propose an algorithm where only the pivot indices $\mathcal{B} = \{J\}_J$ are determined in the decomposition procedure. As contributions from new vectors are subtracted from $\bm{M}$, its diagonal elements decrease monotonously. Consequently, a diagonal $D_p = M_{pp}$ below $\tau$ will never be selected as a pivot element. Since we only determine the pivots, we may screen out elements $M_{pq}$ for which at least one of the corresponding diagonals, ${D}_{p}$ or ${D}_{q}$, is below $\tau$. 
In algorithms where the Cholesky vectors are constructed during the decomposition, screening must instead be with respect to the Cauchy-Schwarz inequality.\cite{koch2003}

Below we outline the procedure to determine $\mathcal{B}$:
\begin{enumerate}
    \item Set $\mathcal{B} = \{\}$.
    \item Determine the significant diagonals:
        \begin{align}
            \mathcal{D}= \{p : D_p \geq \tau \}. \label{eq:standard_screening}
        \end{align}
    For $J \in \mathcal{B}$, only keep $L_{p}^J$ for $p \in \mathcal{D}$. See Fig.~\ref{fig:mem_reduction}.
    \item Find $D_{\text{max}} = \max_{p\in\mathcal{D}}D_p$ and determine the set of qualified diagonal indices $\mathcal{Q}$,
        \begin{align}
            \mathcal{Q}= \{p \in \mathcal{D} : D_p \geq \sigma D_{\text{max}} \},
        \end{align}    
    such that the number of elements in $\mathcal{Q}$ does not exceed $n_{\text{qualified}}^{\text{max}}$. The parameter $\sigma$, which ensures that qualified diagonals are not too small, is called the span factor.
    \item For each $q\in \mathcal{Q}$ compute $M_{pq}$ for all $p\in\mathcal{D}$. If there are any previous Cholesky vectors, subtract their contributions to $\bm{M}$:
        \begin{align}
            \tilde{M}_{pq} = M_{pq} - \sum_{J\in\mathcal{B}}L_p^J L_q^J,\;\; p\in\mathcal{D},\;\; q\in \mathcal{Q}.
        \end{align}
    \item Let $\mathcal{C}$ be the set of qualified indices for which the associated Cholesky vector has been constructed. Initially, $\mathcal{C} = \{\}$.

As long as $\max_{p\in\mathcal{Q}} D_p \geq \tau$, select $q\in\mathcal{Q}$ such that $D_q = \max_{p\in\mathcal{Q}} D_p$, construct the Cholesky vector
        \begin{align}
           L_p^q =\frac{{\tilde M_{pq} - \sum_{J\in\mathcal{C}}L_p^J L_q^J}}{\sqrt{\tilde M_{qq}}}, \;\; p\in\mathcal{D},\;\; q\in \mathcal{Q},
        \end{align}
    update $\mathcal{Q}$ and $\mathcal{C}$,
        \begin{align}
           \mathcal{Q} &= \mathcal{Q} \setminus \{q\},\quad \mathcal{C}= \mathcal{C}\cup \{q\},
        \end{align}
    and the diagonal elements,
    \begin{align}
        D_p = D_p - (L_p^q)^2, \;\; p\in\mathcal{D}.
    \end{align}
    \item Finally, update the pivots $\mathcal{B}$:
        \begin{align}
            \mathcal{B} = \mathcal{B} \cup \mathcal{C}.
        \end{align}
     If $\max_{p\in\mathcal{D}} D_p < \tau$, stop. Otherwise, return to 2.
\end{enumerate}

The memory needed for the Cholesky vectors reaches a maximum during the decomposition and then drops off due to the reduction in the number of elements in $\mathcal{D}$; we only keep $L^J_{p}$ for $p \in \mathcal{D}$. When $\mathcal{B}$ has been determined, $\mathcal{D}$ is empty, and the memory requirement has therefore dropped to zero. 

When $\bm{M}$ is the electron repulsion integral matrix, each pivot $J = \gamma\delta \in \mathcal{B}$ defines a Cholesky basis function $\rho_J(\bm{r}) = \chi_\gamma(\bm{r}) \chi_\delta(\bm{r})$. 
The RI expression,
\begin{align}
L_{\alpha\beta}^J = \sum_{K \in \mathcal{B}} (\alpha \beta | K)Q^{-T}_{KJ}, \quad J \in \mathcal{B}, \label{eq:Cholesky_construct_from_RI_for_us}
\end{align}
may then be used to construct the Cholesky vectors. We decompose $\bm{S}$ and then invert $\bm{Q}$. Note, however, that $\bm{Q}$ may be inverted  while $\bm{S}$ is decomposed.\cite{Benzi1996} 
To approximate $\bm{M}$ to the desired accuracy, we use the Cauchy-Schwarz screening
\begin{align}
    (\alpha\beta | K)^2 \leq (\alpha\beta | \alpha\beta) \cdot  \max_{\gamma\delta} D_{\gamma\delta}  \leq \tau^2. \label{eq:screeningVector}
\end{align}

From the RI formulation, an integral-direct approach is available. By storing $\bm{Q}^{-1}$ and $\mathcal{B}$, the Cholesky vector $L_{\alpha\beta}^J$ may be constructed on-the-fly from Eq.~\eqref{eq:Cholesky_construct_from_RI_for_us}. This may be useful for systems where $\bm{L}$ cannot be stored---the memory required would be proportional to $N_{\mathrm{AO}}^2$ rather than $N_{\mathrm{AO}}^3$.

We use the Libint integral package,\cite{Valeev2017} 
in which $(\alpha \beta | \gamma \delta)$ is computed together with all the integrals in the shell quadruple $(A B | C D)$, where $\alpha \in A$, $\beta \in B$, $\gamma \in C$, and $\delta \in D$. Therefore, the screening and qualification steps are modified such that shell pairs are treated instead of AO pairs. For instance, ${\alpha\beta}\in\mathcal{D}$ if at least one diagonal in $AB$ exceeds $\tau$. 
There is also a trade-off between numerical stability and efficiency: we want to both qualify diagonal indices (add AO pairs to $\mathcal{Q}$) in descending order and compute as few integrals as possible.  Shell pairs $AB$ are therefore ordered with respect to their maximal diagonal element
\begin{align}
D_\mathrm{max}^\mathrm{AB} = \max_{\alpha\beta\in AB} D_{\alpha\beta}.    
\end{align}
Diagonals are then qualified from the $AB$ with the largest diagonal before the next shell pair in the ordered list is considered. To ensure that selected diagonals are not too small, we use $\sigma =10^{-2} $.
In this way, $\mathcal{Q}$ may involve relatively few shell pairs while also containing potential basis elements $J = \alpha \beta$ associated with large diagonals $D_{ \alpha \beta}$.\cite{koch2003}

The procedure described thus far reproduces the integral matrix to within the decomposition threshold $\tau$. However, the framework easily allows for method specific approximations that further reduce the number of elements in $\mathcal{B}$.
We have implemented an active space screening where the target quantities are the molecular orbital (MO) integrals in a selected active space. First, the occupied and virtual AO densities, $\bm{D}^\mathrm{o}$ and $\bm{D}^\mathrm{v} = \bm{S}^{-1} - \bm{D}^\mathrm{o}$, are Cholesky decomposed with the restriction that pivot elements are centered on active atoms. This results in the active occupied density,
\begin{align}
  (\bm{D}^\mathrm{o}_\mathrm{a})_{\alpha\beta} = \sum_{i} C^{\mathrm{a}}_{\alpha i} C^{\mathrm{a}}_{\beta i}, 
\end{align}
 and the active virtual density,
\begin{align}
  (\bm{D}^\mathrm{v}_\mathrm{a})_{\alpha\beta} = \sum_{a} C^{\mathrm{a}}_{\alpha a} C^{\mathrm{a}}_{\beta a}.
\end{align}
The inactive densities are defined as $\bm{D}^\mathrm{o}_\mathrm{i} = \bm{D}^\mathrm{o} - \bm{D}^\mathrm{o}_\mathrm{a} $ and $\bm{D}^\mathrm{v}_\mathrm{i} = \bm{D}^\mathrm{v}_\mathrm{a} - \bm{D}^\mathrm{v} $.\cite{Aquilante2006,SanchezdeMeras2010} To generate the active orbital space, we have adopted the multilevel Hartree-Fock approach of Sæther and coworkers;\cite{Sether2017}  
they use, as $\bm{D}^{\mathrm{o}}$, a superposition of atomic densities\cite{Vanlenthe2006} guess that has been made idempotent by a single Fock matrix diagonalization.
We define the active space screening by replacing the requirements on the diagonals (in steps 2, 3, and 5) with
\begin{align}
   \mathcal{D} = \{\alpha\beta : (\alpha\beta\vert\alpha\beta)v_{\alpha}v_{\beta} \geq \tau\}, \label{eq:alternative_screening}
\end{align}
where
\begin{align}
    v_{\alpha} = \max_p \; (C_{\alpha p}^{\text{a}})^2. \label{eq:orb_screening_vector}
\end{align}
The accuracy of the active MO integrals, rather than the AO integrals, is then controlled by the decomposition threshold $\tau$. The reader is referred to Boman et al.\citep{Boman2008}~for more details on the method specific decomposition approach.

Similarly, only a minor modification of the algorithm is needed to obtain the one-center approximation 1C-CD.
In 1C-CD, the $J = \gamma\delta$ are selected such that $\chi_{\gamma}(\bm{r}) $ and $\chi_{\delta}(\bm{r})$ are centered on the same atom.\cite{Aquilante2007} To implement 1C-CD, we altered the initial screening to exclude all ${\gamma\delta}$ from $\mathcal{D}$ that do not satisfy the one-center requirement.

\section{Results and discussion}

\begin{figure}
    \centering
    \includegraphics[width=\linewidth]{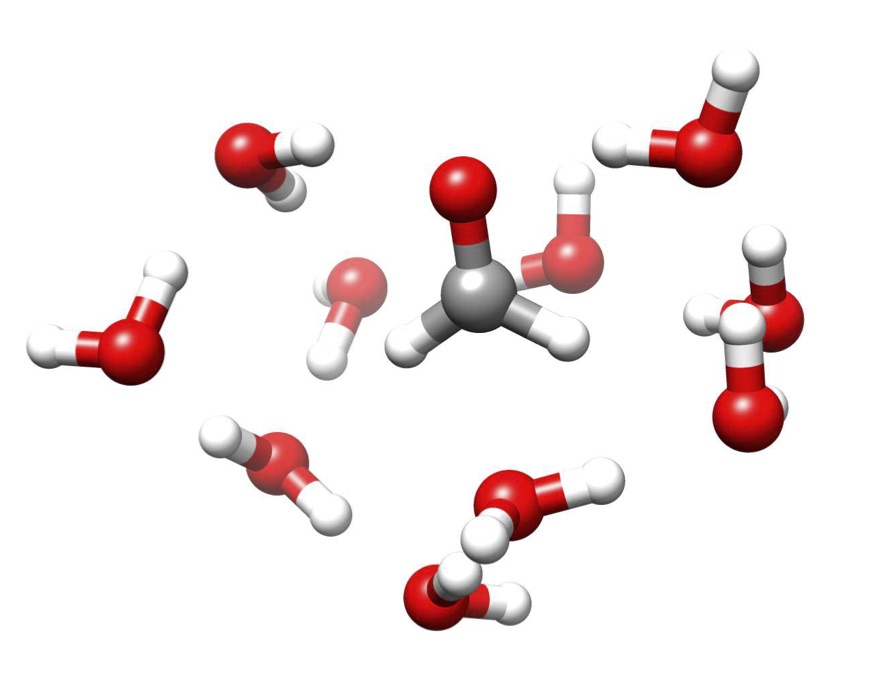}
    \caption{Formaldehyde surrounded by ten water molecules.}
    \label{fig:formaldehyde_10_waters}
\end{figure}

\begin{figure}
    \centering
    \includegraphics[width=\linewidth]{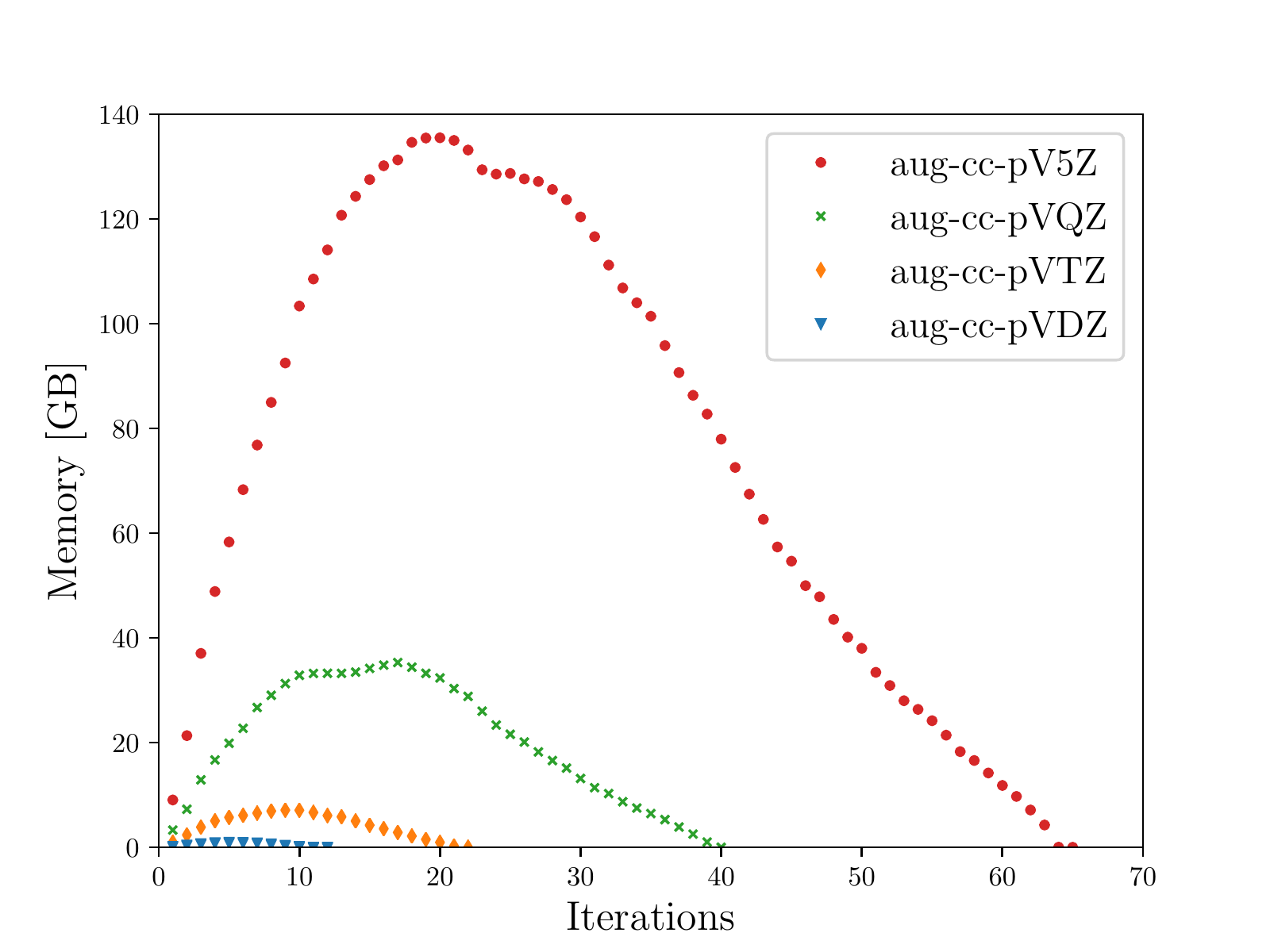}
    \caption{Memory required to hold the Cholesky vectors $\bm{L}$ in each iteration of the decomposition for formaldehyde surrounded by ten water molecules.
    }
    \label{fig:formaldehyde_in_water_mem}
\end{figure}

\begin{table}[hbt]
    \centering
    \caption{Wall time comparisons  between eT and OpenMolcas\cite{molcas} for formaldehyde surrounded by ten water molecules. The total decomposition time is $T = T_1 + T_2$, where $T_1$ and $T_2$ is time to determine $\mathcal{B}$ and $\bm{Q}^{-1}$ and to construct the Cholesky vectors, respectively. Also given is the time to converge the Hartree-Fock equations in QChem,\cite{shao2015} $T_{\text{SCF}}$. Time is in minutes unless other units are specified.  In all calculations, $\tau = 10^{-8}$. Timings were made on an Intel Xeon CPU E5-2699 v4 with 1.5TB shared memory using 22 threads.\label{tab:formaldehyde_10_water}}
    \begin{ruledtabular}
    \begin{tabular}{l c | c c | c c c c | c}
    & & \multicolumn{2}{c|}{OpenMolcas} & \multicolumn{4}{c|}{eT} & QChem\footnotemark[1]\\
    \hline   
    & $N_{\text{AO}}$ & $N_J$ & $T$ & $N_J$  & $T$ & $T_1$ & $T_2$& $T_{\text{SCF}}$ \\
    \hline
   aug-cc-pVDZ & 474  & 5481 & {7} & 5374 & \SI{63}{\second} & 35\si{\second} & 28\si{\second} & \SI{94}{\second}\\
   aug-cc-pVTZ & 1058 & 11184 & 70   & 11212 & 11 & 5 & 6  & 25   \\
   aug-cc-pVQZ & 1972 & 19336 & 589  & 19297 & 79 & 34 & 45 & 249   \\
   aug-cc-pV5Z & 3284 & 30635 & 5534 & 30950 &  498 & 186 & 312 & 7985  \\
     \end{tabular}
    \end{ruledtabular}
    \footnotetext{Version 5.0.2.}
\end{table}
\begin{table}[hbt]
    \centering
    \caption{Cholesky decomposition with $K$ diagonal batches on formaldehyde surrounded by ten water molecules using the aug-cc-pV5Z basis. Here, $N_J$ is the number of Cholesky vectors, $T_1$ the time to determine $\mathcal{B}$ and $\bm{Q}^{-1}$, and $\epsilon$ is the maximal error in the matrix $\bm{M}$. Also given is the peak memory requirement to hold the Cholesky vectors. In all calculations, $\tau = 10^{-8}$.
    }\label{tab:formaldehyde_10_water_partitioning}
    \begin{ruledtabular}
    \begin{tabular}{l c c c c}
   $K$  & $N_J$  & $T_1$ [min] & Memory [GB] & $\epsilon$ \\
    \hline
   1 & 30950 & 186 & 134 & $<\tau$  \\
   2 & 30313 & 158 & 56  & $15\tau$ \\
   4 & 30374 & 123 & 22  & $17\tau$ \\
   6 & 30486 & 106 & 22  & $15\tau$ \\
   8 & 30450 & 90  & 24  & $19\tau$ \\
   10& 30459 & 102 & 25 & $13\tau$ \\
   12& 30407 & 103 & 28  & $16\tau$ \\
     \end{tabular}
    \end{ruledtabular}
\end{table}
\begin{table*}[hbt]
    \centering
    \caption{Full, active space and one-center Cholesky decompositions for the DNA fragment. Here, $N_{\mathrm{AO}}$ is the number of AOs, $\tau$ the decomposition threshold, $N_J$ the number of Cholesky vectors, and $T_1$ is the wall time to determine $\mathcal{B}$ and $\bm{Q}^{-1}$.}\label{tab:dna}
    \begin{ruledtabular}
    \begin{tabular}{l l c  c c c}
    \textbf{Method} & \textbf{Basis} & $N_{\text{AO}}$ & $\tau$ & $N_{\text{J}}$ & $T_1$ [min] \\
    \hline
    \multirow{3}{*}{Full decomposition} &  \multirow{3}{*}{aug-cc-pVDZ}& \multirow{3}{*}{$15064$} & $ 10^{-4}$ & $53742$  & $532$\footnote{Intel Xeon Gold 6132 and 6TB shared memory. Calculation on 140 threads.}\\
    & & & $ 10^{-6}$ & $95403$  & $1854$\footnotemark[1] \\
    & & & $ 10^{-8}$ & $158811$  & $5506$\footnotemark[1] \\
    \hline
    \multirow{2}{*}{Active space decomposition} &{cc-pVDZ/aug-cc-pVTZ}& {$9447$} & {$ 10^{-8}$} & {$19375$}  & {20}\footnote{Intel Xeon CPU E5-2699 v4 and 1.5TB shared memory. Calculation on 22 threads.} \\
    &{aug-cc-pVDZ/aug-cc-pVTZ}& {$15341$} & {$ 10^{-8}$} &{$90551$} & {1389}\footnotemark[1]  \\
    \hline
    \multirow{2}{*}{One-center decomposition} & aug-cc-pVDZ & $ 15064 $ & $ 10^{-8}$ & $ 89489 $ & $ 802$\footnotemark[2]  \\
     & aug-cc-pVDZ & $ 15064 $ & $ 10^{-4}$ & 49533 & 54\footnotemark[2]  \\
     \end{tabular}
     \end{ruledtabular}
\end{table*}
\begin{figure}[htb]
    \centering
    \includegraphics[width=\linewidth]{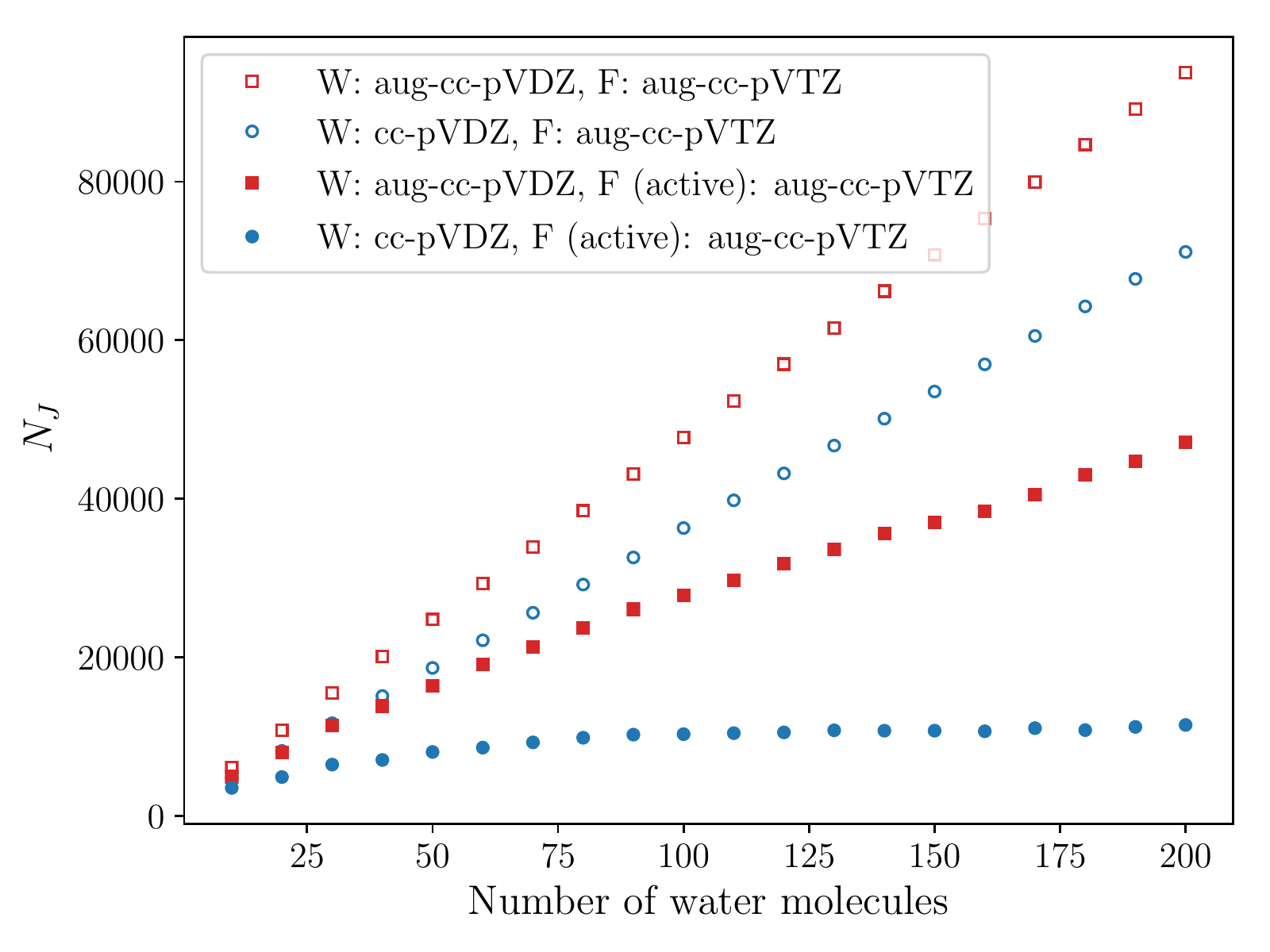}
    \caption{The number of Cholesky vectors $N_J$, in full decomposition and active space decomposition, for formaldehyde (F) surrounded by $10$--$200$ water (W) molecules.}
    \label{fig:formaldehyde_in_water_ml}
\end{figure}
\begin{table*}[hbt!]
    \centering
    \caption{Full, active space and one-center Cholesky decompositions for the retinal-rhodopsin system. Here, $N_{\mathrm{AO}}$ is the number of AOs , $\tau$ is the decomposition threshold, and $N_J$ is the number of Cholesky vectors.}\label{tab:RR}
    \begin{ruledtabular}
    \begin{tabular}{l l c  c c }
    \textbf{Method} & \textbf{Basis} & $N_{\text{AO}}$ & $\tau$ & $N_{\text{J}}$\\
    \hline
    {Full decomposition} &{aug-cc-pVDZ}& { 36787} & {$ 10^{-4}$} & 124632\\
    \hline
    {Active space decomposition} &{cc-pVDZ/aug-cc-pVTZ}& { 23134} & {$ 10^{-8}$} &77719\\
    \hline
   \multirow{4}{*}{One-center decomposition} & cc-pVDZ &  21840 &{$ 10^{-8}$} & 119357 \\
    & aug-cc-pVDZ & 36787 &{$ 10^{-8}$}  & 202935  \\
   & aug-cc-pVDZ & 36787 &{$ 10^{-4}$} &  112592 \\
   & aug-cc-pVTZ & 79420 &{$ 10^{-4}$} &  257198  \\ 
     \end{tabular}
     \end{ruledtabular}
\end{table*}
\begin{figure}[htb]
    \centering
    \includegraphics[width=\linewidth]{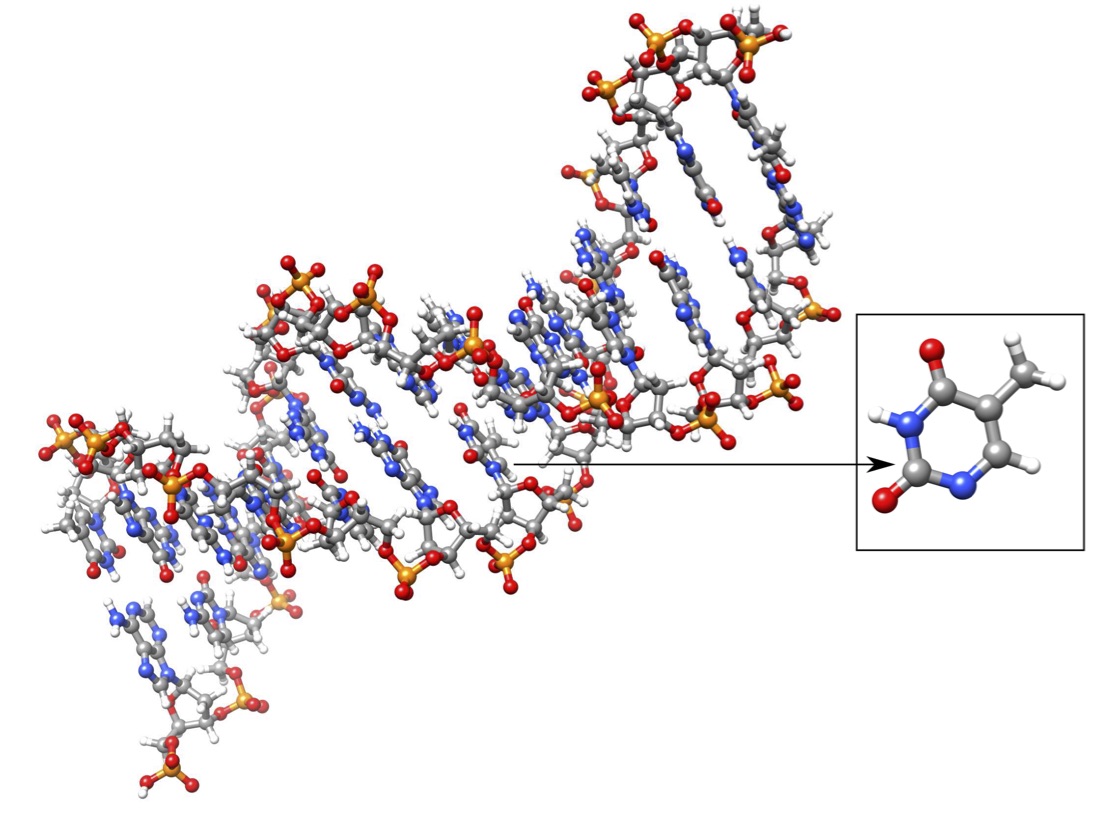}
    \caption{DNA fragment with active thymine.}
    \label{fig:DNA}
\end{figure}
\begin{figure}[htb]
    \centering
    \includegraphics[width=\linewidth]{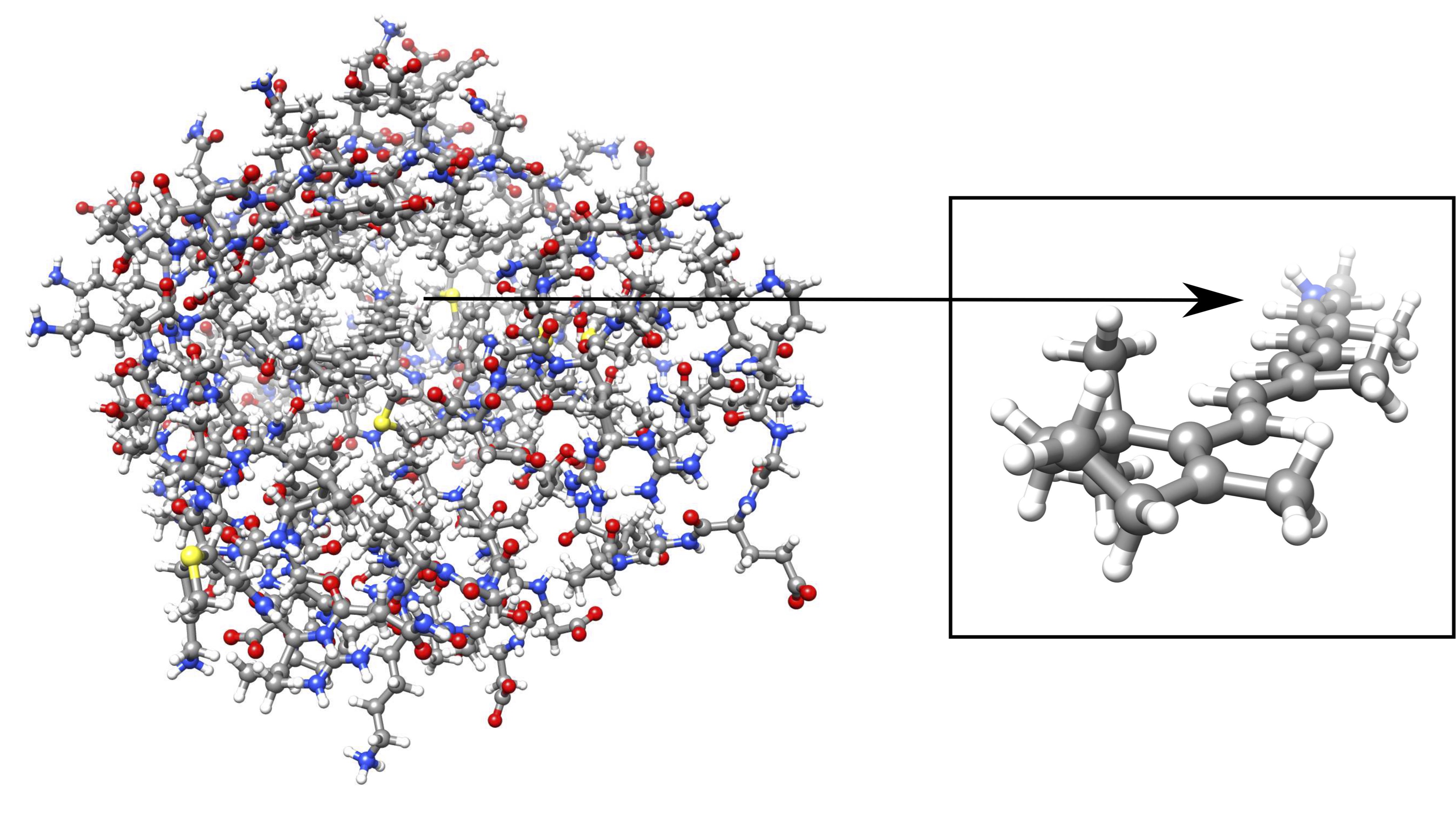}
    \caption{Retinal bound to rhodopsin with active retinal.}
    \label{fig:RR}
\end{figure}

The algorithm was implemented in eT, a coupled cluster program currently under development by the authors and collaborators. To demonstrate its performance, we report wall time comparisons to the OpenMolcas program\citep{molcas} on the formaldehyde-water system in Fig.~\ref{fig:formaldehyde_10_waters}. In these calculations, we use the Dunning basis sets aug-cc-pVXZ, $\mathrm{X} \in \{\mathrm{D}, \mathrm{T}, \mathrm{Q},5\}$.\cite{dunning1989} The results are summarized in Table \ref{tab:formaldehyde_10_water}. Compared to OpenMolcas, the total decomposition time $T$ is reduced by about an order of magnitude. Consequently, as the number of AOs increase, $T$ rapidly becomes negligible compared to the time spent converging the Hartree-Fock equations.

The memory required to hold $\bm{L}$ varies as expected, see Fig.~\ref{fig:formaldehyde_in_water_mem}. It increases to a maximum during the decomposition and then drops off to zero, 
giving a large reduction in memory usage compared to the previous algorithms.\cite{koch2003}
However, to reduce the memory requirements further, the following partitioned matrix algorithm may be used.
First, the significant diagonal is partitioned, $\mathcal{D} = \mathcal{D}_1 \cup \mathcal{D}_2 \cup ... \cup \mathcal{D}_K$, and each diagonal batch decomposed separately, resulting in $\mathcal{B}_1, \mathcal{B}_2, \ldots,$ and $\mathcal{B}_K$. A final decomposition is then performed using $\mathcal{D} = \mathcal{B}_1 \cup \mathcal{B}_2 \cup \ldots \cup \mathcal{B}_K$. With this approach, the decomposition threshold $\tau$ is not an upper bound on the error. However, we have found that the error is controlled by $\tau$ in practice. The error may be lowered by decreasing $\tau$ in all decompositions or only in the final decomposition. We present calculations on the formaldehyde-water system using the aug-cc-pV5Z basis for a set of $K$ values, see Table \ref{tab:formaldehyde_10_water_partitioning}. The peak memory usage is significantly reduced for all $K$ considered, and the time to determine $\mathcal{B}$ and $\bm{Q}^{-1}$ is reduced by up to a factor of two.

Method specific screenings may also be used to treat large systems. Here we apply a multilevel screening, where regions of the system are chosen to be active and the target quantities are the active space MO integrals. We consider an active formaldehyde molecule surrounded by 10--200 water molecules. In Fig.~\ref{fig:formaldehyde_in_water_ml}, we show the number of vectors obtained with the standard and active space screenings defined in Eqs.~\eqref{eq:standard_screening} and \eqref{eq:alternative_screening}.
With the standard screening, the number of Cholesky vectors increases linearly with system size, whereas it flattens out with the active space screening. 
We construct the active orbitals as follows.
The active occupied orbitals are generated from $\bm{D}^{\mathrm{o}}$ by restricting the number of pivots to equal half the number of electrons on the active atoms. In the general case, one pivot is added if an active atom is bound to an inactive atom, effectively adding an orbital to the active occupied space. Similarly, the number of pivots used to decompose $\bm{D}^{\mathrm{v}}$ is restricted such that one obtains the same fraction of virtual to occupied orbitals as in the entire set of orbitals. Alternatively, a decomposition threshold may be used to determine the number of pivots in the decomposition of $\bm{D}^{\mathrm{o}}$ and $\bm{D}^{\mathrm{v}}$.\cite{Sether2017}

The algorithm may be used to decompose the integral matrix of systems with more than ten thousand basis functions. With the method specific and one-center approaches, the applicability of the algorithm is further extended. To show that the algorithm can tackle large systems, we determine $\mathcal{B}$ and $\bm{Q}^{-1}$ for the DNA fragment in Fig.~\ref{fig:DNA}. The time $T_1$ to determine $\mathcal{B}$ and $\bm{Q}^{-1}$,
and $N_J$, are given in Table \ref{tab:dna}. Decompositions using active space screening and the one-center approximation are also listed. 
For the active space calculations, a single thymine is active.

Finally, we present full, active space, and one-center calculations on retinal bound to rhodopsin, see Fig \ref{fig:RR}. Retinal is active in the active space calculations. The number of Cholesky vectors is given in Table \ref{tab:RR}. 

\section{Concluding remarks}
In recent decades, the Cholesky decomposition of the electron repulsion integrals has been implemented in popular quantum chemistry programs. While the technique allows for complete control of the error, a drawback has been its computational cost compared to prefitted RI. With this contribution, the application range of Cholesky decomposition is extended, and its competitiveness with other inner-projection methods improved.
We have already performed full decompositions for systems with tens of thousands of atomic orbitals, yet we expect that the partitioned diagonal approach may be applied to much larger systems. While useful in its own right, the Cholesky decomposition may also be used as an accurate starting point for the development of other integral approximations, such as  the reduced-scaling tensor hypercontraction  schemes.\cite{Hohenstein2012}

\section{Acknowledgements}
We thank A.~S{\'a}nchez de Mer{\'a}s for insightful discussions of the algorithm and its implementation in the early stages of the project.
We also thank E.~Valeev for assistance with the Libint integral package and A.~Krylov for assistance with QChem.

We acknowledge computer resources from NOTUR through project no.~nn2962k and the high performance computer
facilities of the SMART Laboratory.  Furthermore, we acknowledge funding from the Marie Skłodowska-Curie European Training Network through grant agreement no.~765739 (COSINE) and the Norwegian Research Council through FRINATEK project no.~263110/F20.  
H.~K.~acknowledges the Otto Mønsted Fond and S.~D.~F.~acknowledges Fondet til professor Leif Tronstads minne.

\bibliography{./eri_cholesky.bib}

\end{document}